\documentclass[10pt,conference]{IEEEtran}
\IEEEoverridecommandlockouts
\usepackage{cite}
\usepackage{amsmath,amssymb,amsfonts}
\usepackage{algorithmic}
\usepackage{graphicx}
\usepackage{textcomp}
\usepackage{xcolor}

\usepackage{url}

\usepackage{booktabs} 

\usepackage{graphicx}
\usepackage{caption}
\usepackage{subcaption}
\usepackage{amsmath}
\usepackage{balance} 
\usepackage{amssymb}
\usepackage{xspace}
\usepackage[T1]{fontenc}
\usepackage[scaled=0.81]{beramono}
\usepackage{balance}
\let\labelindent\relax
\usepackage{enumitem}  
\usepackage{listings}
\usepackage{url}
\usepackage[ruled,lined,linesnumbered,vlined]{algorithm2e}
\usepackage{multirow}
\usepackage{array}
\usepackage{color}
\usepackage{float}
\usepackage[misc]{ifsym}

\newcommand{\tool}{\textsc{EasyFlow }}
\newcommand{\ethereum}{Ethereum }
\newcommand{\myparagraph}[1]{\vspace*{0.14cm}\textbf{#1.}\quad}
\newcommand{\etal}{{\emph{et al.}} }

\def\BibTeX{{\rm B\kern-.05em{\sc i\kern-.025em b}\kern-.08em
		T\kern-.1667em\lower.7ex\hbox{E}\kern-.125emX}}
\begin{document}
	
\title{\tool: Keep Ethereum Away From Overflow
	 \thanks{This paper has been published in Proceedings of ICSE 2019 \copyright IEEE }
}

\author{
	\IEEEauthorblockN{Jianbo Gao}
	\IEEEauthorblockA{School of EECS\\
		Peking University\\
		Beijing, China\\
		gaojianbo@pku.edu.cn}
	\and
	\IEEEauthorblockN{Han Liu}
	\IEEEauthorblockA{School of Software\\
		Tsinghua University\\
		Beijing, China\\
		liuhan2017@tsinghua.edu.cn}
	\and
	\IEEEauthorblockN{Chao Liu}
	\IEEEauthorblockA{School of EECS\\
		Peking University\\
		Beijing, China\\
		liuchao\_cs@pku.edu.cn}
	\and
	\IEEEauthorblockN{Qingshan Li}
	\IEEEauthorblockA{School of EECS\\
		Peking University\\
		Beijing, China\\
		liqs@pku.edu.cn}
	\and
	\IEEEauthorblockN{Zhi Guan \Letter}
	\IEEEauthorblockA{National Engineering Research Center for Software Engineering\\
		Peking University\\
		Beijing, China\\
		guan@pku.edu.cn}
	\and
	\IEEEauthorblockN{Zhong Chen}
	\IEEEauthorblockA{School of EECS\\
		Peking University\\
		Beijing, China\\
		zhongchen@pku.edu.cn}
}


\maketitle

\begin{abstract}
While \ethereum smart contracts enabled a wide range of 
blockchain applications, they are extremely vulnerable to 
different forms of security attacks. Due to the fact that 
transactions to smart contracts commonly involve cryptocurrency 
transfer, any successful attacks can lead to money loss or even 
financial disorder. In this paper, we focus on the overflow attacks in 
\ethereum, mainly because they widely rooted in many smart contracts 
and comparatively easy to exploit. 
We have developed \tool, 
an overflow detector at \ethereum Virtual Machine level. 
The key insight behind \tool is a taint analysis based tracking 
technique to analyze the propagation of involved taints.
Specifically, \tool can not only divide smart contracts into 
safe contracts, manifested overflows, well-protected overflows and potential overflows, 
but also automatically generate transactions to trigger potential overflows.
In our preliminary evaluation, 
\tool managed to find potentially vulnerable \ethereum contracts with little 
runtime overhead. A demo video of \tool is at \url{https://youtu.be/QbUJkQI0L6o}.
\end{abstract}


\section{Introduction}
\label{sec:intro}
Blockchain and smart contracts have been widely applied and adopted
since the decentralized cryptocurrency Bitcoin was first introduced by Nakamoto\cite{nakamoto2008bitcoin}.
Ethereum provides a quasi-Turing-complete virtual machine\cite{buterin2013ethereum} and allows developers to program smart contracts to implement complex functions.
Developers can even issue a new cryptocurrency
based on Ethereum with only several hundred lines of source code under an ERC Token Standard.
However, smart contracts are vulnerable 
and hackers are intensely attracted as they are closely related to money.
Overflow vulnerabilities of smart contracts are especially easy to be exploited
compared to different forms of security attacks.

In April 2018, 
BecToken was attacked by integer overflow on multiplication, 
causing an extremely large amount of tokens transferred to malicious accounts and the price of BEC cleared.
The vulnerable function is located in batchTransfer and the code is shown in Fig.~\ref{fig:example-bec}.
As indicated in line 4,
the second parameter \texttt{\_value} can be any 256-bit integer and the result of multiplying \texttt{\_value} by \texttt{cnt} may overflow.
In this particular transaction, 
\texttt{\_value} was set to 0x8000...000 (63 0s) and \texttt{\_receivers} were two different addresses, thus the result overflowed and \texttt{amount} was calculated to be 0.
Each of these two receivers received about 5.79E58 BECs  but 0 BEC was deducted from the sender's account.

\begin{figure}[h]
\begin{lstlisting}[frame=none, numbers=left, language=Java, showstringspaces=false, morekeywords={function,contract,uint,mapping,address,event,bytes,uint256}]
function batchTransfer(address[] _receivers, uint256 
       _value) public whenNotPaused returns (bool) {
  uint cnt = _receivers.length;
  uint256 amount = uint256(cnt) * _value;
  require(cnt > 0 && cnt <= 20);
  require(_value>0 && balances[msg.sender]>=amount);
  ...
  return true;
}
\end{lstlisting}
\caption{A vulnerable function of \texttt{BEC}}
\label{fig:example-bec}
\end{figure}

Although overflow has become one of the most devastating vulnerabilities,
there are few effective detection schemes and tools.
From the practical perspective, the main challenges of detecting overflow vulnerabilities are as follows.

\myparagraph{Challenge 1: Infer and Trigger Potential Overflows}
Overflow only occurs in certain transactions 
with special input data and message value. 
Traditional testing techniques are insufficient to 
infer and trigger potential overflow vulnerabilities in attack-free transactions 
and have difficulty generating test cases 
because StateDB must be considered in the \ethereum setting,
which consists of account addresses, balances, global variable values.

\myparagraph{Challenge 2: Identify Protection Patterns}
Experienced developers may implement effective protection schemes
such as SafeMath library and assertions to protect their contracts from overflow,
which may lead to a significant increase in false positive in the detection tools. 

\myparagraph{Our Insight}
To address the challenges, we developed \tool 
for detecting overflow vulnerabilities in smart contracts.
The key insight behind \tool is a taint analysis based tracking 
technique to analyze the propagation of involved taints.
In the detection, \tool monitors the transaction process,
captures manifested overflows, identifies well-protected overflows,
and automatically generates transactions to trigger potential overflows.
We have applied \tool to detect real transactions on \ethereum Mainnet,
and found it efficient to discover vulnerable contracts even from attack-free transactions.

\section{Overview}
\label{sec:framework}

\begin{figure}[H]
\centering
\includegraphics[width=8.5cm]{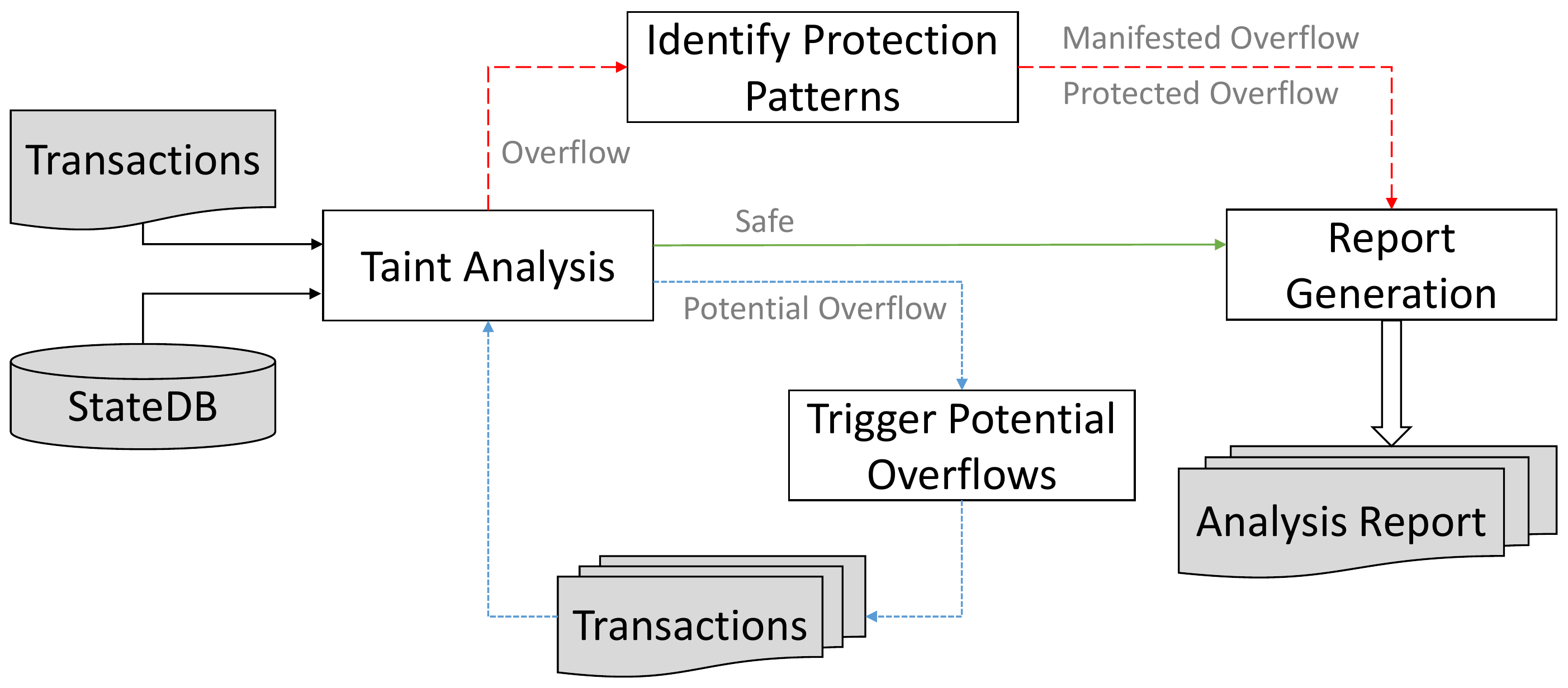}
\caption{The high-level framework of \tool.}
\label{figure:framework}
\end{figure}

The general work flow of \tool is shown in Fig.~\ref{figure:framework}. 
Specifically, \tool consists of four components in high-level design.
It takes transactions as input, 
and StateDB is accessible which includes key-value pairs in storage, balances of accounts and codes of smart contracts.

Taint Analysis component tracks taints inside the EVM interpreter 
to monitor the transaction process.
The transactions are first analyzed by Taint Analysis component
and the smart contracts are divided into three categories: safe, overflow and potential overflow.
Report Generation will output analysis reports for safe contracts directly.
\tool will try to identify protection patterns in overflows,
and refine them into manifested overflows and protected overflows.
When potential overflows are inferred, 
\tool will generate a series of transactions
in order to trigger the potential overflows via re-execution. 
As long as any one of the generated transactions is manifested overflow,
it means the original potential overflow has been triggered.


\subsection{Taint Analysis at EVM Level}
\label{sec:taint_analysis}
\tool extends the EVM interpreter by tracking the propagation of taints during transactions execution. 
Taking the \texttt{batchTransfer} function in Fig.~\ref{fig:example-bec} as an example again,
the taints are introduced when the parameters \texttt{\_receivers} and \texttt{\_value} are loaded.
When \tool executes instructions corresponding to the source code in line 4,
the values and taint marks of \texttt{cnt} and \texttt{\_value} are moved to the top of stack and \texttt{MUL} is then executed to calculate the value of \texttt{amount}. 
As \texttt{MUL} is a susceptible integer arithmetic instruction,
\tool will check both the marks and values involved in this instruction.
This instruction is safe if the two multipliers are not taints
and the protection patterns will be further analyzed when the multiplication result overflows.
If more than one multipliers is tainted but the multiplication result does not overflow, \tool infers that it is a potential overflow and try to trigger it later.

\subsection{Identify Protection Patterns}
\label{sec:handle_protection}
Experienced developers have proposed some schemes to protect their smart contracts from overflow attacks, leading to high false positives in detection tools.
The most commonly used library for protection is named SafeMath\footnote{\url{https://openzeppelin.org/api/docs/math_SafeMath.html}}  which provides several functions for developers to replace ordinary arithmetic operations in contracts.
The protection scheme behind SafeMath is using \texttt{require} or \texttt{assert} function to check the result after calculation and the contract will throw when an overflow occurs.
Some variant libraries and protection codes are also widely adopted and have similar logics and patterns to SafeMath.
Therefore, it is not difficult to enumerate all the patterns
of protection schemes. 
 

\subsection{Trigger Potential Overflows}
\label{sec:generate_transaction}
An \ethereum transaction takes input data and message value as external data and all the internal data is stored in StateDB.
Input data is the payload that the sender sends in a transaction to call a specific function in a smart contract and passes in parameters, and message value indicates how many ETHs are sent.
The format of input data is a 32-bit function signature and several 256-bit integers, each of which represents a parameter, the position of an array or the length of an array.

To trigger an inferred potential overflow,
we implement a straight-forward algorithm in \tool.
The first generated transaction has the same input data as the original transaction but \texttt{MAX\_UINT256} as message value. 
Then the 256-bit integers in input data are split and each is assigned to 0 and \texttt{MAX\_UINT256}.
\tool takes all the possible combinations of these integers as input data and the original message value to generate transactions for re-execution.

\section{Using \tool}
\label{sec:usage}
\begin{figure}[H]
\centering
\includegraphics[width=8.5cm]{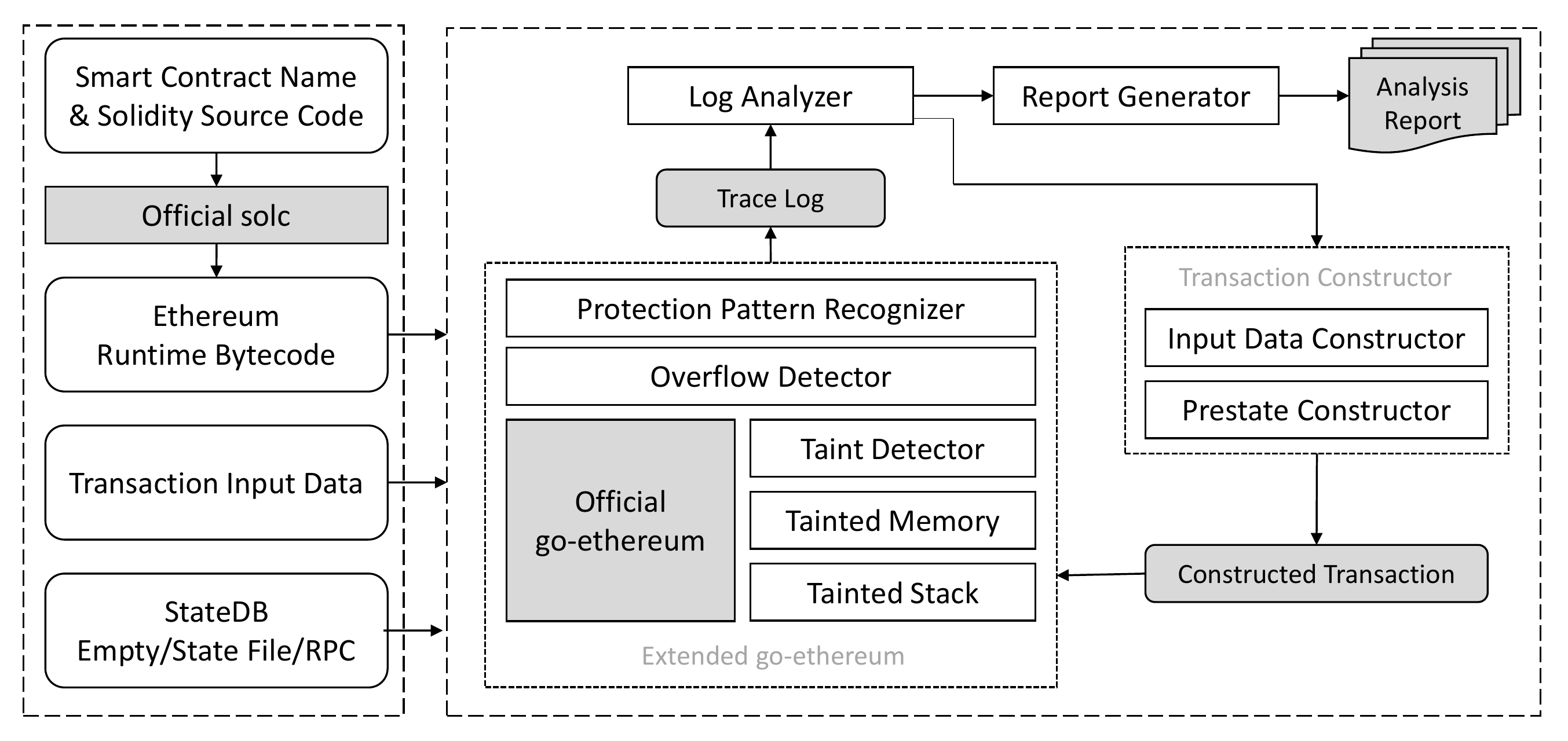}
\caption{The architecture of \tool in implementation}
\label{figure:architecture}
\end{figure}

As Fig.~\ref{figure:architecture} illustrates, 
\tool can be divided into four parts in implementation,
extended go-ethereum, log analyzer, transaction constructor and report generator.
Ethereum runtime bytecode and transaction input data can be passed into \tool,
and Solidity source code of smart contracts will be compiled 
using a built-in official solc in advance. 
The default StateDB is empty, and state information can be passed in via a JSON-format file before detection, or a running \ethereum node can be connected via remote procedure call (RPC) to provide the real-time StateDB.

\myparagraph{Extended go-ethereum}
Extended go-ethereum was developed 
based on official golang implementation of the Ethereum protocol, 
aiming at tracing the propagation of taint data.
Tainted stack and tainted memory 
followed Ethereum stack and memory in implementation,
but recoding taint marks instead of specific values while EVM running.
Taint detector can mark input data as a taint 
via inspecting instructions and analyzing their intents, 
Overflow detector will check both taint marks and values
when EVM executes susceptible integer arithmetic instructions,
such as \texttt{ADD}, \texttt{SUB}, \texttt{MUL} and \texttt{EXP}.
To reduce false positive, 
in case of smart contract developers using SafeMath library 
or conditional statements to protect contracts from overflow,
pattern recognizer will try to confirm 
whether it is a manifested overflow or a protected overflow
by matching bytecode to protection patterns.
Extended go-ethereum will output trace logs in JSON format
as soon as finishing dynamic bytecode execution.

\myparagraph{Log Analyzer}
Log Analyzer analyzes logs generated by extended go-ethereum,
and distributes tasks to corresponding components.
Log parser first extracts the overflow detection result from logs.
Safe, manifested overflow, protected overflow 
and re-executed potential overflow transactions 
will be sent to report generator.
Transaction constructor will receive potential overflow transactions
to conduct further analysis.

\myparagraph{Transaction Constructor}
Transaction Constructor is implemented for 
re-executing potential overflow transactions with constructed input data.
As indicated in \ref{sec:generate_transaction}, 
Transaction Constructor splits input data, 
combines new values of input data and message value,
and constructs new transactions.
After construction, the transactions 
will be re-executed by extended go-ethereum 
to discover whether the potential overflow vulnerability can be triggered.

\myparagraph{Report Generator}
Report generator gathers every received result,
and extracts key information into a brief analysis report.
All the log files are also attached, 
and can be accessed through links in report.

\begin{figure}[H]
\centering
\includegraphics[width=.85\linewidth]{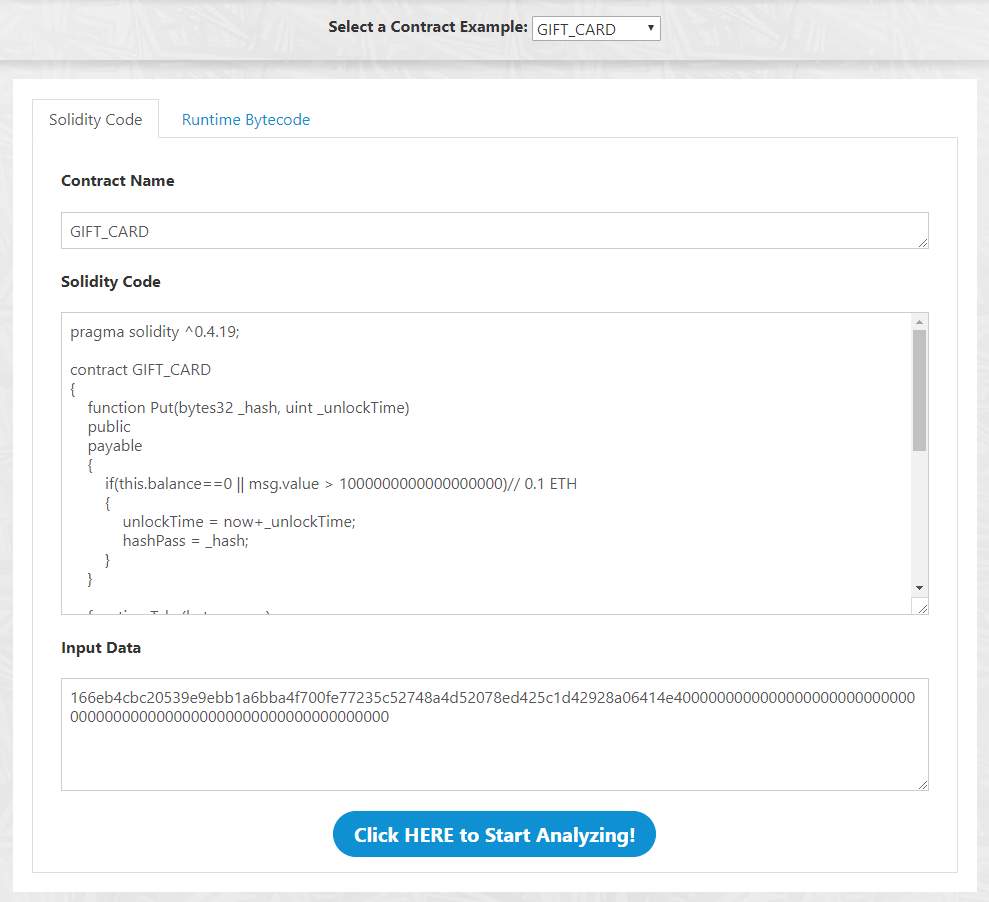}
\caption{\tool Screenshot}
\label{figure:screenshot}
\end{figure}

We have deployed \tool as a web service\footnote{\url{http://easyflow.cc/}},
and a command-line tool having the same functions 
can be used offline on Linux as well.
Fig.~\ref{figure:screenshot} shows
a screenshot of the web page of \tool.
\tool provided ten smart contract examples 
containing contract name, Solidity source code, runtime bytecode and input data.
By clicking the button below,
\tool will automatically analyze 
overflow vulnerabilities of the selected smart contract,
and generate a succinct analysis report at the bottom.
Users can also modify contents of the examples 
or type brand new contract code and input data into the web page,
and get the specific analysis report with "one-click".

The report shown in Fig.~\ref{figure:screenshot} consists of analysis result and analysis log.
A sentence of conclusion is used as the analysis result,
and input data and result of all the executed transactions 
are included in the analysis report. 
Users can also download full trace logs of each transaction 
via the link on the right of the transaction number. 
All the log files is in JSON format 
and can be easily read by machine for further processing.

\begin{figure}[H]
\centering
\includegraphics[width=8.5cm]{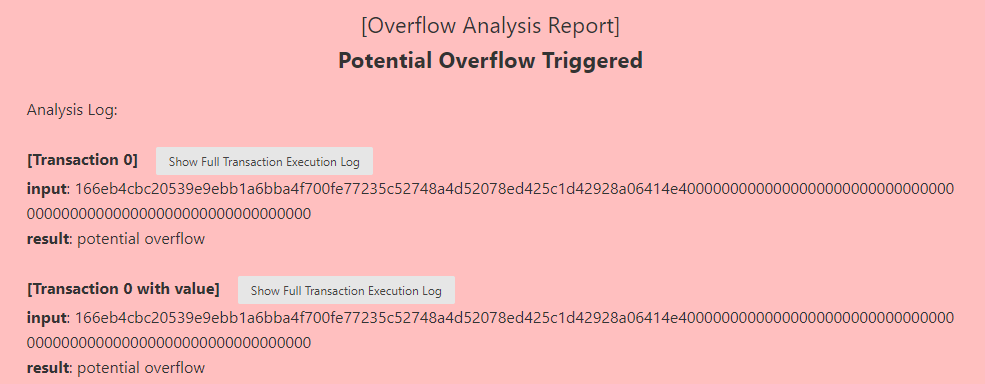}
\caption{\tool Analysis Report}
\label{figure:report}
\end{figure}

\section{Preliminary Evaluation}
\label{sec:eval}
We have preliminarily evaluated \tool on \ethereum Mainnet
and the detection results in TABLE~\ref{table:result} show that \tool is effective in detecting overflow vulnerabilities in smart contracts and can successfully identify protection patterns and trigger potential overflows.

\begin{table}[!hbt]  
\centering  
\caption{Detection results} 
\begin{tabular}{rr}  
  \hline  
  \textbf{Detection Result}&\textbf{Number of Txs} \\
  \hline
  Manifested Overflow&465\\
  Protected Overflow&6\\
  Potential Overflow Triggered&3871\\
  Potential Overflow Not Triggered&42580\\
  Safe&19914\\
  \hline
\end{tabular}
\label{table:result}
\end{table} 

All the experiments were conducted on Amazon Web Services Cloud (AWS). 
The virtual machine that deploys \tool has 1 vCPU, 512MB RAM, 
and Ubuntu Server 16.04 LTS - Xenial (HVM) as the OS.
An \ethereum full node provides StateDB via RPC and it is running on a 
virtual machine having 4 vCPUs, 32GB RAM, SSD, and the same OS.
For most of the smart contracts,
\tool finished analysis in 300ms in command-line mode, 
and the execution time increases linearly compared to official go-ethereum.

\newcommand{\tabincell}[2]{
  \begin{tabular}
  {@{}#1@{}}#2
  \end{tabular}
}
\begin{table}[!hbt]  
\centering  
\caption{Analysis results on example smart contracts. MO: manifested overflow. 
PO: protected overflow. POT: potential overflow but triggered. PONT: potential overlow 
not triggered. S: safe.} 
\begin{tabular}{rcccccr}  
  \hline  
  \textbf{Contract}&\textbf{MO}&\textbf{PO}&\textbf{POT}&\textbf{PONT}&\textbf{S}&\textbf{Overhead} \\
  \hline
  BecToken&$\checkmark$&&&&&104.35\%\\
  
  SMT&$\checkmark$&&&&&98.15\%\\

  Lizun&&$\checkmark$&&&&100.00\%\\

  darx&&$\checkmark$&&&&103.45\%\\

  Gift\_Card&&&$\checkmark$&&&496.15\%\\

  RedEnvelope&&&$\checkmark$&&&294.34\%\\

  SimpleLotto&&&&$\checkmark$&&750.00\%\\

  HBToken&&&&$\checkmark$&&631.03\%\\

  Rating&&&&&$\checkmark$&92.31\%\\

  Danksignals&&&&&$\checkmark$&110.71\%\\
  \hline
\end{tabular}
\label{table:eval}
\end{table} 

The analysis results of some example smart contracts 
are shown in TABLE~\ref{table:eval}
and the addresses of the smart contracts and transactions 
can be accessed on Github\footnote{\url{https://github.com/Jianbo-Gao/EasyFlow/tree/master/taint-realworld}}.
BecToken is the most famous contract having an overflow vulnerability, 
and was exploited because of unprotected multiplication.
SMT has the similar vulnerability because of unprotected addition. 
Lizun and darx are protected from overflow vulnerability
using SafeMath Library and assertions.
GIFT\_CARD is a simple addition overflow, 
and \tool successfully triggered the vulnerability 
by automatically constructing input data based on the real-world transaction.
The triggered overflow vulnerability of RedEnvelope is caused by message value.
It cannot be exploited in the real world at this time 
as the amount of ETH is limited,
but may be exploitable with the increase of ETH.
SimpleLotto actually has an overflow vulnerability,
but the contract had committed suicide and all the state of this contract was cleared
so that the vulnerability cannot be triggered without manually constructing state information.
The potential subtraction overflow in HBToken is protected by conditional statements, 
and can never be triggered by any input data.
There are not susceptible instructions
in the transactions of Rating and DANKSIGNALS,
and they are both considered to be safe.

\section{Related Work}
\label{sec:rw}
Research on integer overflow has been in progress for decades
and many effective detection schemes were proposed.
RICH\cite{brumley2007rich}, BRICK\cite{chen2009brick},  SmartFuzz\cite{molnar2009dynamic} and AIR\cite{dannenberg2010if} 
were developed to detect integer overflows.
Dietz \etal\cite{dietz2012understanding} focused on the integer overflows in C and C++ code.
Sun \etal\cite{sun2016inteq} showed that not all the integer overflows were malicious
and it could be checked using equivalence checking across multiple precisions.

As blockchain becomes popular in both academia and industry,
the smart contract bug detection are noted by researchers.
Oyente\cite{luu2016making} is a symbolic execution tool 
built to find potential security bugs of Ethereum contracts.
ZEUS\cite{kalra2018zeus} presents a formal verification framework 
that can build and verify correctness and fairness policies.
Some other schemes and tools \cite{liu2018reguard}\cite{liu2018s} are also presented to
analyze smart contracts for vulnerabilities.
In contrast, \tool focus more on integer overflow vulnerabilities,
and it is capable of detecting various types of overflows in real-world contracts 
via dynamic taint analysis, pattern matching and transaction construction.

\section{Conclusion}
\label{sec:conclusion}
In this demo paper, we present \tool, a virtual machine level 
detector for overflow vulnerabilities in Ethereum. The key insight 
behind \tool is a taint analysis based tracking technique to monitor 
real transactions. Particularly, \tool captures manifested overflows, flags 
well-protected overflows, infers and triggers potential overflows as well. 
We managed to leverage \tool to find real overflows in deployed 
smart contracts. In the future, we plan to generalize the technique 
to diverse settings and applications.

\section*{Acknowledgment}
This work is supported by National Natural Science Foundation of China under the grant No.: 61672060, 61802223 and China Postdoctoral Science Foundation under Grant No.: 2017M620785.

\linespread{0.9}	
\bibliographystyle{IEEEtran}
\bibliography{taint}

\end{document}